\documentclass[sigconf,authorversion,nonacm,screen]{acmart}
\AtBeginDocument{%
  \providecommand\BibTeX{{%
    \normalfont B\kern-0.5em{\scshape i\kern-0.25em b}\kern-0.8em\TeX}}}

\setcopyright{acmlicensed}
\copyrightyear{2024}
\acmYear{2024}
\setcopyright{rightsretained}
\acmConference[CHI EA '24]{Extended Abstracts of the CHI Conference on Human Factors in Computing Systems}{May 11--16, 2024}{Honolulu, HI, USA}
\acmBooktitle{Extended Abstracts of the CHI Conference on Human Factors in Computing Systems (CHI EA '24), May 11--16, 2024, Honolulu, HI, USA}
\acmDOI{10.1145/3613905.3650918}
\acmISBN{979-8-4007-0331-7/24/05}

\usepackage{CJKutf8}
\usepackage{soul}
\usepackage[vskip=1em,font=itshape,leftmargin=2em,rightmargin=2em]{quoting}

\usepackage{subfigure}

\makeatletter
\def\@AtEndDocument{\relax}
\makeatother

\begin{document}
\begin{CJK}{UTF8}{ipxm}

% Thesis title: Influence of positive initial engagement on the acceptability of embodied conversational agents (ECAs) for older adults
% Positive First Impressions of an Embodied Conversational Agent Unrelated to Acceptance by Older Adults
\title[No Joke]{No Joke: An Embodied Conversational Agent Greeting Older Adults with Humour or a Smile Unrelated to Initial Acceptance}

\author{Ge ``Rikaku'' Li}
\orcid{0009-0001-9588-7294}
\affiliation{%
  \institution{Tokyo Institute of Technology}
  \city{Tokyo}
  \country{Japan}
}
\email{li.g.ai@m.titech.ac.jp}

\author{Katie Seaborn}
\email{seaborn.k.aa@m.titech.ac.jp}
\orcid{0000-0002-7812-9096}
\affiliation{%
  \institution{Tokyo Institute of Technology}
  \city{Tokyo}
  \country{Japan}
}

\renewcommand{\shortauthors}{Li and Seaborn}

\begin{abstract}
  Embodied conversation agents (ECAs) are increasingly being developed for older adults as assistants or companions. Older adults may not be familiar with ECAs, influencing uptake and acceptability. First impressions can correlate strongly with subsequent judgments, even of computer agents, and could influence acceptance. Using the circumplex model of affect, we developed three versions of an ECA---laughing, smiling, and neutral in expression---to evaluate how positive first impressions affect acceptance. Results from 249 older adults indicated no statistically significant effects except for general attitudes towards technology and intelligent agents. This questions the potential of laughter, jokes, puns, and smiles as a method of initial engagement for older adults.
\end{abstract}




\begin{teaserfigure}
  \includegraphics[width=\textwidth]{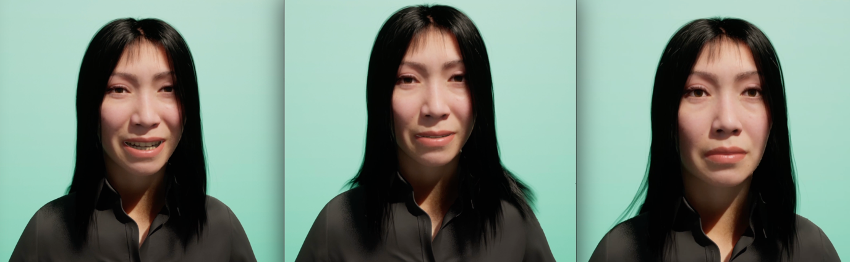}
  \caption{Three versions of the ECA, each representing a different affective state based on varying levels of valence and arousal: laughing (left), smiling (middle), neutral (right).}
  \Description{Three images of the same ECA are presented side-by-side, from the shoulders up. The first is laughing. The second is smiling. The third is neutral. The ECA has the appearance of an Asian women with short dark hair and a dark dress shirt.}
  \label{fig:teaser}
\end{teaserfigure}


\maketitle

\section{Introduction}

Embodied conversational agents (ECAs) are interactive computer systems that mimic real agents---typically people, but also non-humans---in appearance and/or behaviour~\cite{olson2003human,ruttkay2004brows}. The visual appearance of ECAs compared to other socially intelligent agents (IAs) allows them to communicate with users non-verbally. Nonverbal behaviors, such as nods, posture, and facial expressions, play an important role in engagement~\cite{ishii2020impact}, trust and empathy~\cite{yalccin2018modeling}, and emotional expression~\cite{ahuja2019react,mcneill1992hand}, among people and with computer agents~\cite{calisgan2012identifying}.
ECAs are increasingly being explored for older adults as emotional support~\cite{loveys2022felt}, everyday support~\cite{smrke2022aging}, and for healthcare~\cite{ter2020design}. The realism of ECAs can be attractive~\cite{yalccin2018modeling} but also uncomfortable, if ``too'' humanlike or uncanny~\cite{thaler2020agent}. Also, older adults experience more barriers to acceptance of new technologies compared to other age groups~\cite{ashjian2019implementation,etemad2019senior,gitlow2014technology,hauk2018ready}. Therefore, ECA design must be carefully geared towards older adults. %and that can positively influence their interaction with the ECA. 

In human-human interaction, first impressions are crucial and can be lasting~\cite{bar2006very,leary2011personality,knapp2007handbook}---and the same appears to be true when it comes to computer agents, like ECAs~\cite{binsted1995using,olson2003human}. %Leary and Alle [46] did a social psychology research which showed that personality for human plays an important role in the way people manage how they convey themselves through self-presentations, trying to affect the audience's first impressions and increase effectiveness. Olson and Olson [67] showed that rules related to face-to-face communication can be extended to Human-Computer Interaction (HCI) contexts, such as with ECAs. 
This may be especially true for older adults, when first impressions are formed based on an agent's nonverbal expressions of personality and interpersonal attitudes~\cite{cafaro2016first}. %In the first user-agent contact, specific nonverbal immediacy cues, such as smiling, gazing, and proximity behaviors influence the user's interpretation of agent extraversion and subordination and affect the user's relational decisions regarding the likelihood and frequency of further contact. Nonverbal behaviors may influence the first impressions older adults have towards ECAs as well, including long-term perceptions of certain ECAs or ECAs in general, but this is far less unexplored.
One nonverbal characteristic that could influence first impressions and help older adults embrace ECAs more readily is humour~\cite{kulms2014let,nijholt2002embodied,cann2001perceived}. Indeed, affective computing researchers have called for further study on ECA emotional expression and humour~\cite{nijholt2002embodied}. %If a person is perceived to have a sense of humour, this assumption automatically has a halo effect on other desirable personality traits such as friendliness, cheerfulness, and creativity [11]. 
At present, very little research has been conducted on the impact of humour in ECAs generally or specifically within the context of first impressions and for older adults. %Although cognitive decline associated with aging can affect aspects of humour comprehension and the ability to be humourous, understanding humour is still generally less affected by age, and jokes can even be perceived as funnier by older adults than other age groups [28]. However, this has not yet been explored for ECAs with older adults.

As a first step, we evaluated whether acceptance of an ECA by older adults could be influenced through first impressions featuring humour. %, expressed verbally and non-verbally. 
We asked: \textbf{\emph{Can humour improve initial acceptance of an ECA by older adults by influencing first impressions?}} To this end, we built an ECA with three distinguishable levels of affective expression that introduces itself via video. We randomly assigned older adults to each version of the ECA. We found no statistically significant differences, except in relation to general technology and IA acceptance.
%by level of familiarity with technology or IAs, nor by ECA. %I grounded this work in the Senior Technology Acceptance Model (STAM), which is based on the Technology Acceptance Model (TAM) [14, 40]. 
We report our negative results in accordance with human-computer interaction (HCI) and general science procedures against publication biases~\cite{cockburn2020threats,mlinaric2017dealing}. We contribute these null findings and raise questions about the who, where, and how of humour for ECA design.
%Our study addresses gap in research on ECAs with humour in older adults and provide guidance for the design of ECAs for older adults.

%% ----------------- BACKGROUND -----------------
\section{Theoretical Background}

\subsection{Older Adults, Agent Acceptance, and Humour}
Populations around the world are ageing at a rapid pace~\cite{united2017department}.
%, and Japan is no exception. According to the Statistics Bureau of Japan [90], the proportion of older adults (65 years old and above) in Japan will be almost 30% by 2022. 
Advances in everyday technologies parallel this trend, with older adults emerging as a key user group. 
%Therefore, “older adults and technology” has become a popular research and development topic.
Anderson and Perrin~\cite{anderson2017technology} found that in 2016, 67\% of older adults used the Internet and about 40\% had a smartphone. %This compares to 12% and 18% in 2000, indicating that older adults are also moving toward a more digital lifestyle. 
Still, technology adoption among older adults is low compared to 90\% of the general adult population being ``wired in.'' %Although Internet use among older adults is gradually increasing, a number of challenges and barriers remain. These barriers include the complexity of technical equipment, security issues, language barriers, and barriers to the physical condition of older adults [20]. Moreover, 
Older adults experience barriers, e.g., complex and inaccessible user interfaces (UI)~\cite{diehl2022perceptions}, and may also resist new technology, which is linked to low acceptance~\cite{vaportzis2017older,hanson2010influencing}.

%2.3	Agent Embodiment for Older Adults: Voice and Body
Autonomous IA with multimodal embodiments, like ECAs, may be an inclusive and engaging option~\cite{von2010doesn,el2018towards,simpson2020daisy}. %Voice-based agents, in particular, can increase perceptions of friendliness [17] and social presence [76]. 
Still, first impressions are critical, perhaps especially for older adults~\cite{karahanouglu2011perceived}. By ``first impressions,'' we mean the initial feelings and attitudes---positive, negative, and neutral---elicited in a first encounter~\cite{bar2006very}. First impressions are formed unconsciously and quickly, usually within a few seconds. Importantly, if the first use is negative, subsequent use may be disrupted or perceived as negative~\cite{gardner1993computer}. Facial expressions, such as smiling, can lead to positive first impressions~\cite{lee2019study,nunamaker2011embodied}. Humour may be especially effective~\cite{kuhnlenz2013increasing,ring2013addressing,ter2020you,binsted1995using}. For instance, Tabak et al.~\cite{ter2020you} found that older adults appreciated dynamic elements such as humour over static characteristics such as agent agedness and genderedness. 
Binsted at el.~\cite{binsted1995using} suggested that self-deprecating humour could be employed by an agent. As such, we designed our ECA to express humour, including self-deprecating humour, verbally (with voice) and nonverbally (with facial expressions).

\subsection{Modeling Technology Acceptance}
Davis~\cite{davis1989perceived} used the Theory of Rational Action (TRA) to infer that beliefs about technology can affect intention to adopt technology: the Technology Acceptance Model (TAM)~\cite{davis1989perceived}. Core concepts include perceived ease of use and perceived usefulness. TAM predicts that people are more likely to adopt a technology when it is perceived to be easy to use and useful, and perceived ease of use can influence perceived usefulness~\cite{davis1989perceived}. %Scholars have extended the model by adding other contextual variables, including social and environmental influences [86].
%TAM is a useful theoretical framework for investigating the determinants of internal use beliefs. Although a large body of scholarly work has focused on the adoption of Computer-Mediated Communication (CMC) technologies such as email, telecommunications, the Internet, and e-commerce, scholars have extended TAM to the context of HCI. 
For example, Lee at el.~\cite{Lee2020} explored whether and to what degree older adults would accept a soft service robot in the home, finding that perceived ease of use, usefulness, and subjective norms were statistically significant predictors of acceptance. %However, they also found that perceived anxiety and perceived favourability pairs were not significantly predictive with respect to the willingness to adopt and use soft service robots for older adults.
Chen at el.~\cite{Chen2014} later developed the Senior Technology Acceptance Model (STAM) using the TAM and other models, including the Unified Theory of Acceptance and Use of Technology (UTAUT)~\cite{venkatesh2012consumer}. The elder-centred STAM includes eight predictors: geriatric technology self-efficacy, geriatric technology anxiety, facilitation conditions, self-reported health status, cognitive ability, social relationships, life attitudes and satisfaction, and physical functioning. %) after controlling for socio-demographic factors such as e.g., age, gender, education level, and economic status. Geriatric technology is about an electronic or digital product or service that increases the independent living and social participation of older adults in relative health, comfort, and safety.
Acceptance of geriatric technology is operationalized as positive attitudes towards technology. 
%Geriatric technology self-efficacy is the user's assessment of their ability to successfully perform tasks using geriatric technology, while geriatric technology anxiety refers to the emotional responses triggered by using geriatric technology to perform tasks, such as worry, nervousness, or anxiety. Facilitating conditions are environmental factors that help older adults use geriatric technology more easily. The authors tested STAM in 1,012 older adults in Hong Kong, and their results support the proposed model. China and Japan share similar cultural roots and technology upkeep. Since I am looking at older adults in Japan, I used the STAM as the underlying theoretical model, notably in the questionnaire design and interpretation of the results.
%In summary, the acceptance of ECAs for older adults is important. ECAs have the potential to deliver care benefits in supporting people affected by dementia and their caregivers, and they could be an important tool in improving the quality of life for older adults. Understanding the factors that influence the acceptance of ECAs among older adults can help in designing more effective and user-friendly ECAs that can improve the quality of life for older adults. 
As such, an ECA that introduces itself in a positive way may influence older adults' first impressions in a similarly positive way, and thus their acceptance of the ECA.

Combining the potential of a humorous ECA and the STAM, we hypothesized that: \textbf{\emph{H1. Use of humour expressed by an ECA through nonverbal facial expressions and verbal content, i.e., laughing, puns, and jokes, will improve its acceptance by older adults.}} We compared this version to a smiling one (a low arousal complement) and one with a neutral expression. We used measures from STAM research. Since socio-demographic factors, including level of familiarity with technology, could have an effect~\cite{Chen2014}, we also compared novice and non-novice groups.

%% ----------------- METHODS -----------------
\section{Methods}
We conducted a between-subjects online\footnote{We used SurveyMonkey: \url{https://www.surveymonkey.com/}} experiment, where each group viewed a greeting video by the ECA with one of three expressions: laughing with a pun and joke, smiling with a pun, and neutral with a pun (refer to \autoref{sec:materials} and \autoref{fig:teaser} for details). 
Our protocol was registered before data collection\footnote{Registered on April 2\textsuperscript{nd}, 2023, at OSF: \url{https://osf.io/yhtd6}}
% https://osf.io/9a7wf/?view_only=c565c01d899144ae8264ed1498771fe4
and approved by the IRB (\#2023064).
%2023064.

\subsection{Participants}
%Before the recruitment, I calculated the sample size by using G*power, results showed that I needed at least 53 people in each group. After recruitment, 
Japanese older adults (N=249; women n=60 or 24\% and men n=189 or 76\%, of which n=2 were X-gender\footnote{X-gender is similar to non-binary in Japanese culture.} and n=2 declined to report details) aged 65--80~\cite{united2017department} (M = 69.6, SD = 3.6) were recruited through Yahoo! Crowdsourcing\footnote{Yahoo! JAPAN ensures unique respondents and quality through identity verification: \url{https://crowdsourcing.yahoo.co.jp/}} between June 12\textsuperscript{th} and June 13\textsuperscript{th}, 2023. We excluded records less than 6 minutes, the bare minimum time according to video length and pilot tests.
%where the timestamp was less than 900 seconds because the videos add up to more than 180 seconds, and it took Japanese people about 600 seconds to go through the video and the questionnaire, so the results of these questionnaires were likely to be inaccurate.
Older adults were pseudo-randomly assigned to ECA group by birthday; this and data quality eliminations led to different totals per group: n=78 in laughing, n=69 in smiling, and n=102 in neutral.
% By group: n=43 in laughing, n=38 in smiling, and n=59 in neutral.
%The general health status across the 3 groups was as follows (average score): Laughing group (M = 3.2, SD = 0.7, Smiling group (M = 3.5, SD = 0.7), Normal group (M = 3.3, SD = 0.5). There was a significant difference between Laughing group and other groups (p < 0.005). Even so, all groups were slightly good, in terms of health status overall. 
76\% (n=190) had never seen or heard about ECAs before, and
96\% (n=240) had never used ECAs before.
% 85.7\% (n=120) had never seen or heard about ECAs.
%This indicated that for most participants, this was the first time that they had seen or heard an ECA. Therefore,  did they think about ECAs was based on their first impressions.
Participants were paid in accordance with the participant pool at roughly 1200 yen/hour, equating to 400 yen for 25 minutes.

\subsection{Procedure}
Participants answered the general attitudes questionnaire (refer to \autoref{sec:measures}). Then they watched the ECA self-introduction video (refer to \autoref{sec:materials}) and answered an attention check question (input the random number at the end of the video). They also answered questions about their attitudes towards the ECA (refer to \autoref{sec:measures}). %The videos displayed by YouTube. 
Finally, they provided demographics and received a Yahoo! Crowdsourcing code for compensation.

\subsection{Measures and Instruments}
\label{sec:measures}

All responses were rated on a 5-point Likert scale ranging from 1 (strongly disagree) to 5 (strongly agree), unless noted. 

\subsubsection{Technology Acceptance (GAT, GAIA)}
We used the nine TAM items from the STAM model-based~\cite{Chen2020}, validated 14-item instrument by Chen at el.~\cite{Chen2014} to capture general attitudes towards technology (GAT) and IA specifically (GAIA). Content included: attitude towards use, perceived usefulness, perceived ease of use, technology anxiety, technology self-efficiency, facilitating conditions, self-reported conditions, social relationships, attitudes towards ageing, and life satisfaction. We excluded the health and capability items, since the ECA context of use was not medical.

\subsubsection{Acceptance of the ECA (AECA)}
We followed the measurement selection protocol of Kramer at el.~\cite{Kramer2022}, using the same measures with 26 items for acceptance based on use of the ECA: relationship with ECA, usability, enjoyment, aesthetics, privacy concerns, control, and perceived usefulness. Items on direct interaction were excluded because we used videos for first impressions. %Therefore, I did not use any irrelevant measures (a 10-item rapport scale [26], Creating Rapport with Virtual Agents [33], Classic aesthetics [45], Concern for privacy scale [13], Active control [53], Perceived usefulness scale [19]).

%% KS: WERE THESE NOT USED OR ANALYZED?????
%\subsubsection{Relationship with ECA (Rapport Scale)}
%Gratch et al. [27] explored the phenomenon of rapport, a feeling of relatedness generated by quick and occasional positive feedback between people and the virtual character, often associated with socio-emotional processes. They used a questionnaire to measure users' subjective sense of rapport with ECA. I choose only the questions that matched the research design and presentation of the ECA in my experiment, i.e., video-based self-introductions, because there was no direct interaction with the ECA. The response was on a 5-point Likert scale ranging from 1 (strongly disagree) to 5 (strongly agree).:
%“Did you feel you had a connection with the other person?”
%3.4.2.2	Enjoyment
%Most existing studies of Internet use are based on the Technology Acceptance Model (TAM) and therefore did not include two particularly important constructs, namely, convenience and social factors. In this research, Cheung at el. [16] adopted and extended the Triandis model to form a new theoretical model for studying Internet/WWW usage acceptance. All responses were on a 5-point Likert scale ranging from 1 to 5:
%3.4.2.3	Perceived Usefulness
%Heerink et al. [33] proposed a model of technology acceptance based on Unified Theory of Acceptance and Use of Technology (UTAUT), which was used to measure the technology acceptance of older adults. All responses used a 5-point Likert scale ranging from 1 (strongly disagree) to 5 (strongly agree). The model is shown in Table 2.

\subsubsection{Helpfulness Potential (HP) and Helpfulness Ratio (HR)}
Behavioural responses are difficult to measure in online experiments, but procedures may be adapted from previous research. Porath and Erez~\cite{Porath2007} considered how rudeness affects task performance and desire to help, which they operationalized as the average number of pencils that each group helped the researchers pick up. %The average for the neutral group was ~8 (89.8%), while the rude group picked up ~2 (35.5%), indicating that people who experienced rudeness were less willing to help. 
Kühnlenz at el.~\cite{kuhnlenz2013increasing}, at the end of the experiment, gave participants the option to go directly to the final stage or to help the robot with an object labeling task. Agreeing to this additional, arduous task was used to measure a basic willingness to help the robot, and the number of labeled pictures produced was used as an indicator of the degree of help. %Subjects were faced with a rather long list of pictures and could leave at any time after the first five labeled pictures.
%The results showed that those who were not positively adapted to the robot chose to escape and did not help the robot at all, linking positive affect and helpfulness behaviour. Another identical peak can be observed from 21 to 60 labeled pictures, where 8 subjects in both groups stopped helping the robot, while some pictures started repeating the first repeated picture. Nevertheless, some participants (5 in the implicitly emotionally adapted group and 4 in the non-adapted group) continued to help the robot mark up to 71-80 pictures, which is almost half of the subjects who showed the maximum amount of help in the emotionally adapted group. Inferring from the significant group differences in picture labeling, participants facing full emotional adaptation showed higher helpfulness to the robot than those in the non-adaptive group. Therefore, it may a useful tool to measure helpfulness. 
In Wood et al.~\cite{Wood2013}, word count was used to measure children's willingness to talk to a robot by comparing the total number of words children used in their interactions with the robot versus in the interactions with a human interviewer, helping the researcher to understand the differences in their performance in these two contexts. This method provided a quantitative way to compare children's interactions in different interviewer contexts by translating verbalizations into numerical data.

Based on the above, we created the Helpfulness Potential (HP) and Helpfulness Ratio (HR) measures to test willingness to help the ECA. %During the study, it was imperative to ensure that the chosen tasks were closely aligned with the ECA's embodiment, i.e., being a conversational, algorithm-driven agent, and the study's design, especially the online context. Given that real-world interactions with ECAs predominantly involve language-based communication rather than physical actions, I aimed to select tasks that resonated with this fundamental aspect. Compared to other tasks, such as “help pick up the pen,” which may not mirror the predominant language-based interactions users have with ECAs, 
Given the ``conversational'' context, we asked participants to imagine a conversation with the ECA. We captured (i) the percentage of those willing to help train the ECA, or HP (similar to picking up a pen~\cite{Porath2007} or an object labeling task~\cite{kuhnlenz2013increasing}) and (ii) how many words participants were willing to contribute, or HR (like dialog word counts between children and a robot~\cite{Wood2013} and how many labels were produced to help a robot~\cite{kuhnlenz2013increasing}).

%The ECA performed a self-introduction via video in the questionnaire, so the participants did not actually have direct interaction with the ECA. Still, this allowed participants to develop a first impression of the ECA by observing the ECA's emotional state without any other factors interfering (with the exception of the participants who had seen or heard of the ECA before). Based on this first impression, I asked participants to write a scripted conversation with the ECA for training the ECA to perform better. By analyzing the participants' willingness to engage in this task and the number of words they used, it was possible to observe their willingness to contribute to the improvement of the ECA's performance. I provided an “easy means of escape” like Kühnlenz at el. [41], where participants could easily skip this procedure or choose to help the ECA, increasing the credibility of the study. Participants were unaware of the actual purpose of the study and an objective and quantitative rubric was used, increasing accuracy of this measure.

The task proceeded as follows. First, participants were given the option to help or not help train the ECA (HR):

\begin{quoting}[indentfirst=false]
Thank you for taking the time to let us know what you think of our agents. We appreciate your help in training our agents to speak better. (エージェントの感想をお聞かせいただき、ありがとうございます。エージェントをもっとうまく話せるように訓練するために、ご協力をお願いいたします。)\\
A.	Yes, I'll help (はい、お手伝いします)\\
B.	I'll just finish the last questions (結構です、最後の質問を終わらせます)
\end{quoting}

Then they were asked to write a conversational script between themselves and the ECA (HP) using this template: %I did not include the template parts, e.g., みかん：, in the word count. (This part is for measuring the HP).

\begin{quoting}[indentfirst=false]
Please use the following template: (テンプレートに関しては、以下のものをご使用ください)\\
Me (私)：\\
Mikan (みかん)：\\
Me (私)：\\
Mikan (みかん)：\\
(...)
\end{quoting}

\subsubsection{Demographics}
Demographics included age, gender, health condition, familiarity with ECAs (seen or heard before, or not), and, if applicable, use frequency of ECAs and which kinds used.

\subsection{Materials}
\label{sec:materials}

\subsubsection{Design of the ECA}
We created three versions of the ECA using Unreal Engine 5, Live Link Face 1.2.1, and Metahuman Creator\footnote{\url{https://metahuman.unrealengine.com/}} by Epic Games. %In the production of ECAs, I divided the production process into two parts, the first part was to create the appearance of ECAs, this process used Metahuman Creator. 
Based on Kulms at el.~\cite{kulms2014let} and our Japanese population, we chose a feminine Asian avatar in formal dress.
%looked at the effect of gender on partner perceptions and found that ECAs with female gender were rated higher in positive partner perceptions than male ECAs in a companionship context, therefore I created the image of adult female with a formal shirt. 
Animating the agent in line with the voice and script  %involved controlling the ECA so that the mouth-lip movements and speech would correspond and provide rich expressions when the ECAs spoke. For this process, I
involved capturing a Japanese lab member's facial expressions and lip movements on a mobile device (iPhone 11) using Live Link Face and transferring these to Unreal Engine to sync with Metahuman Creator.
%Next, I recorded three self-introduction videos (Laughing, Smiling and Neutral) with different facial expression and jokes. In Figure 3, 
We used the mappings of Lim and Aylett~\cite{lim2007new}, based on the circumplex model of affect~\cite{russell1980circumplex}, to identify three distinct facial expressions and how to animate these using the ECA's facial features: laughing for the humour condition (high arousal, positive valence), smiling as a non-humorous complement (low arousal, positive valence), and neutral as a control (low arousal, neutral valence). The lab member first studied Figure 9 from Lim and Aylett~\cite{lim2007new}, focusing on the eyebrows, eyes, and mouth of the faces. Then, they imitated each expressions while reading the associated script aloud.
%Valence values from 0 to 0.5 indicate negative prices and valence values greater than 0.5 indicate positive prices. A shift in valence value from negative to positive will move the lip curvature from a downward U to an upward U, as shown in Figure 5. In extremely happy situations, the cheek bulge is visible below the eyes. In contrast, in extremely unpleasant situations, wrinkles formed next to the flanks of each nostril due to the nasolabial folds. In the arousal dimension, the size of the open eyes increased with increasing arousal and decreased with decreasing arousal. The eyebrows are affected by both arousal and valence value. In the positive effect state, when arousal is low to medium (less than 0.5), the eyebrows show a slight V-curve. As arousal increases, the brow becomes more relaxed and straight. When arousal is very high (over 0.8), the eyebrows are raised slightly, and the raised inner eyebrows form subtle wrinkles on the forehead. The values of 0.5 and 0.8 can basically be considered as watersheds for different facial expressions and are very important. Therefore, the facial expression I made for the ECA of Laughing one looks like to the face in the upper right corner, and the arousal and valence of the facial expression of Smiling one is between 0.6-0.8, and the arousal and valence of the facial expression of Smiling one is between 0.4-0.6.

The script was a self-introduction. For all conditions, we set a baseline positive tone using a pun related to the ECA's name (``Although I`m not an edible `Mikan,' I absolutely love oranges!''). For the ``laughing'' condition, we used a joke related to technology\footnote{\url{https://911cybersecurity.com/tech-jokes-a-collection-of-computer-network-infrastructure-and-cybersecurity-humor}} that a Japanese lab member translated and back-translated (``Do you know why the computer is feeling cold? Ha-ha! It is because Windows is left open!''). The ECA introduced its name, functions, and interests (refer to Appendix A in Supplementary Materials). All videos are on OSF~\footnote{\url{https://osf.io/9a7wf/files/osfstorage}}, with links to each video listed in Appendix B (Supplementary Materials).

\subsubsection{Pilot Tests and Manipulation Checks}
We conducted three pilot tests to ensure (i) the design of the ECA%and choices of emotional expressions were accurate
, i.e., manipulation checks, (ii) no deficiencies in the survey, e.g., bugs, typos, Japanese language checks, and (iii) no instances of bias. Eight Japanese lab members (1 woman, 7 men) participated in all tests.
%1 woman and 7 men, all Japanese, participated in all pilot tests.
%The process of the pilot test was almost same as the real experiment. Firstly, I displayed the three videos to 8 lab members (1 woman, 7 men) at a random order, then they need to finish the questionnaire. Finally, I talked about the whole process of the virtual experiment and the design of the ECA.
%3.5.2.1	Pilot Test for Confirming the Design of the ECA’s Emotional Levels
The first focused on the three emotional expressions: whether these matched the expected arousal and valence levels and were distinguishable. Each person voted on the level of valence and arousal for each ECA. The sums matched expectations; however, all thought that the laughing ECA was a bit unnatural. As such, a hand was added, in line with social norms for women laughing in Japanese culture.
%3.5.2.2	Pilot Test for Checking the Robustness of the Questionnaire
The second pilot test focused on the content and logic of the three questionnaires (one for each version of the ECA). We modified the base questionnaire by adding explanations about the ECA at the start and editorial fixes.
The third pilot test focused on the new ECA design for the laughing condition. The original version and two new versions with a hand over mouth or under chin were evaluated based on arousal and valence. Results indicated that the ``hand under chin'' version was highest in valence, so we used this video for the laughing condition in the study.

% \begin{figure}[htbp]
%   \centering
%   \includegraphics[width=\textwidth]{Figs/Fig-Pilot3.png}
%   \caption{Options for the laughing ECA following the first pilot test.}
%   \Description{Two options are presented, showing the ECA laughing with (i) hand over chest and (ii) hand over mouth.}
%   \label{fig:pilot1}
% \end{figure}

\subsection{Data Analysis}
\label{sec:analysis}

Descriptive statistics were generated for all measures (GAT, GAIA and AECA) by ECA version and self-reported familiarity with ECAs: novice (no knowledge) and non-novice (some knowledge or more). Shapiro-Wilk tests were used to determine normality. Most were atypical, so non-parametric tests, e.g., Kruskal-Wallis, were used for analyzing the differences between groups.
%and Kendall's tau-b tests for analyzing correlations. And I used 
GoTranscript\footnote{\url{https://gotranscript.com/}} was used for calculating the number of Japanese words for HP.

%% ----------------- RESULTS -----------------
\section{Results}
\label{sec:results}

\begin{figure*}[ht]
  \centering

    \subfigure[Helpfulness potential (HP) results.]{%
        \label{fig:results-hp}
        \includegraphics[width=0.45\textwidth]{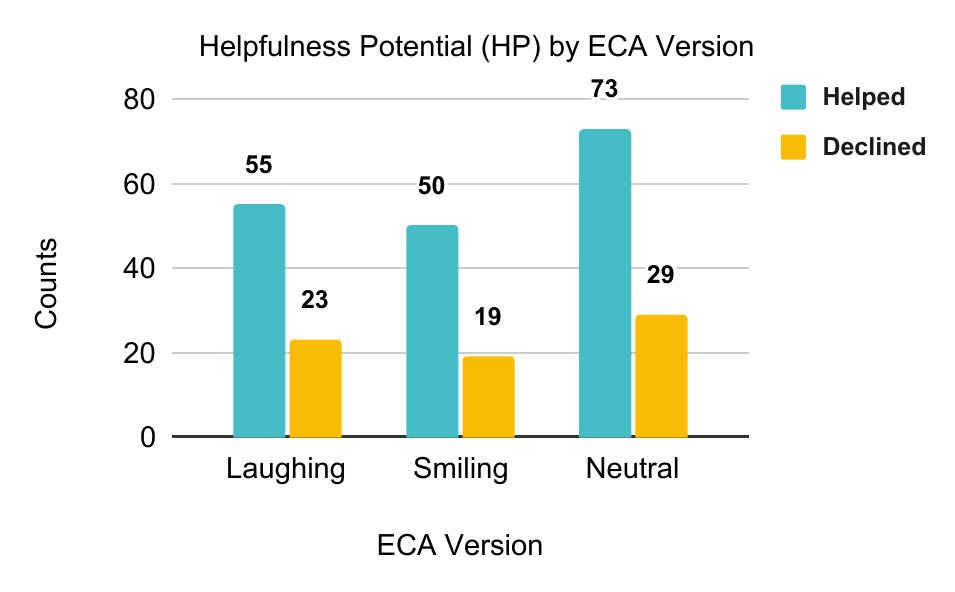}
    }%
    \subfigure[Ratio for HP results.]{%
       \label{fig:results-hpratio}
       \includegraphics[width=0.45\textwidth]{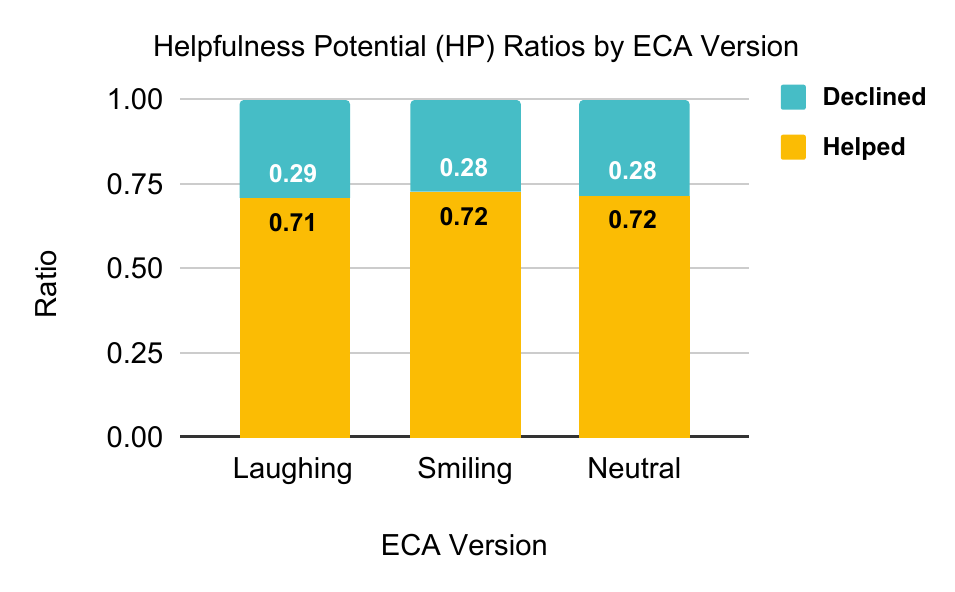}
    }\\ %  ------- End of the first row ----------------------%
    \subfigure[Helpfulness ratio (HR) results.]{%
        \label{fig:results-hr}
        \includegraphics[width=0.45\textwidth]{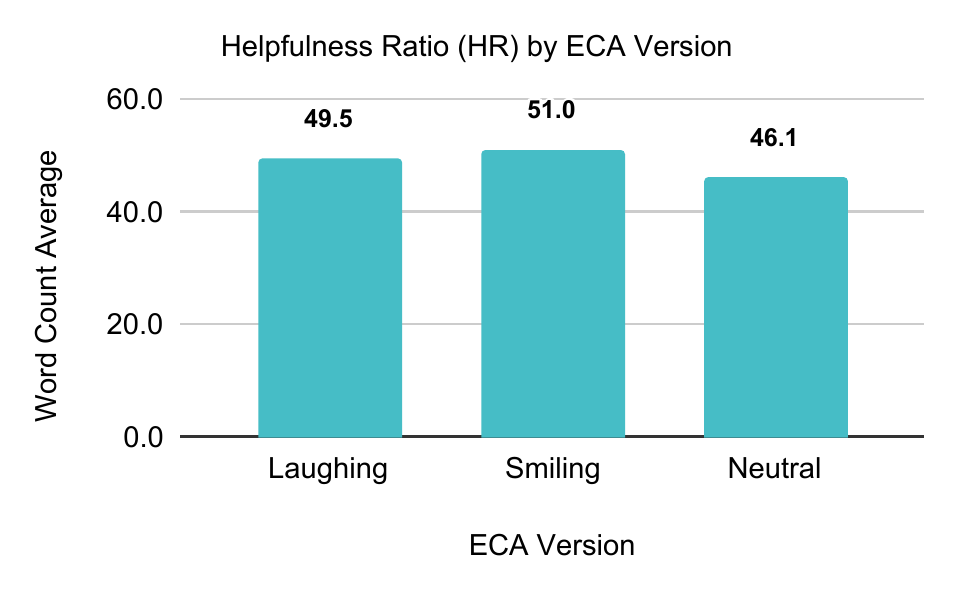}
    }%
    \subfigure[Acceptance of the ECA (AECA) results.]{%
        \label{fig:results-aeca}
        \includegraphics[width=0.45\textwidth]{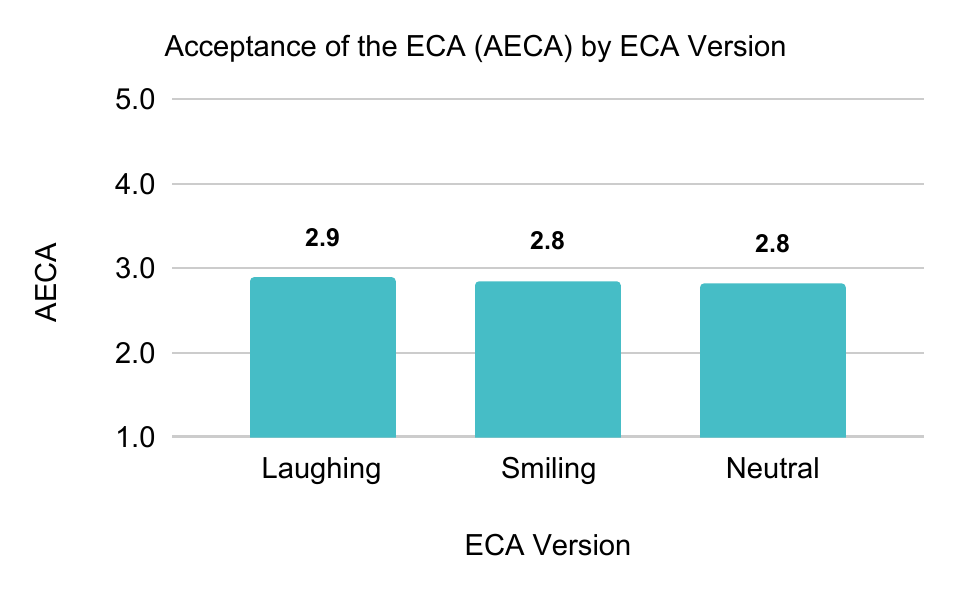}
    }%

    \caption{Key results for behavioural and attitudinal measures by ECA version.}
    \Description{Four graphs showing the behavioural results, notably the helpfulness potential or HP and the helpfulness ratio or HR, as well as the acceptance of the ECA or AECA results.}
  \label{fig:results}
\end{figure*}

We present our results below, with key results provided in \autoref{fig:results}.

\subsection{General Acceptance of Technology (GAT) and Intelligent Agents (GAIA)}

Descriptive statistics for the GAT were: laughing group (M = 3.3, SD = 0.6), smiling group (M = 3.4, SD = 0.5), and neutral group (M = 3.4, SD = 0.6). For the GAIA: laughing group (M = 3.3, SD = 0.6), smiling group (M = 3.3, SD = 0.6), and neutral group (M = 3.3, SD = 0.5). Kruskal–Wallis tests found no statistically significant difference by ECA group for GAT (\emph{p} = .842) and GAIA (\emph{p} = .890). A Mann-Whitney U test found a statistically significant difference in GAT between novices (M = 3.3, SD = 0.6) and non-novices (M = 3.6, M = 0.4), \emph{U} = 2400.5, \emph{Z} = -2.61, \emph{p} < .001, but not for GAIA, \emph{p} = .096. This indicates similar levels of technology and IA acceptance for all older adults, but flags a potential difference between novices and those more familiar with general technology.

\subsection{Acceptance of the ECA (AECA)}

A Kruskal-Wallis test found no statistically significant difference between the laughing (M = 2.9, SD = 0.6), smiling (M = 2.8, SD = 0.6), and neutral (M = 2.8, SD = 0.6) groups, \emph{p} = .420 (Figure~\autoref{fig:results-aeca}). This suggests that first impressions, regardless of emotional expression, did not have an impact on acceptance of the ECA. 
%Descriptive statistics for AECA were: laughing group (M = 3.0, SD = 0.4), smiling group (M = 3.0, SD = 0.3), neutral group (M = 2.8, SD = 0.4). 
A Mann-Whitney U test also did not find a statistically significant difference for AECA between novices (M=2.8, SD=0.6) and non-novices (M=3, SD=0.5), \emph{p} = .196, meaning older adults accepted the ECA regardless of general familiarity with ECAs. However, a Spearman's rank correlation test for GAT and AECA was statistically significant, \emph{r}(247) = .422, \emph{p} < .001, as was one for GAIA and AECA, \emph{r}(247) = .454, \emph{p} < .001. Older adults who were more open to accepting technology or IA in general were also more accepting of our ECA. In sum, general acceptance over first impressions.

%For ECA novices: laughing group (M = 3.0, SD = 0.4), smiling group (M = 3.0, SD = 0.3), neutral group (M = 2.9, SD = 0.4). For those with ECA experience: laughing group (M = 3.2, SD = 0.6), Smiling group (M = 2.8, SD = 0.3), neutral group (M = 2.8, SD = 0.4).

%A Pearson correlation test found a statistically significant correlation between GAT and AECA, \emph{r}(138) = .415, \emph{p} < .001, as well as GAIA and AECA, \emph{r}(138) = .448, \emph{p} < .001. An ANOVA found no statistically significant difference between ECA novices and users (\emph{p} = .072, \emph{p} = .214, \emph{p} = .075). In short, if older adults were open to accepting technology or IA in general, then they were more accepting of our ECA. Even so, Mann–Whitney U tests were run to compare each version of the ECA, but no statistically significant differences were found for AECA. 

\subsection{Helpfulness Potential (HP) and Helpfulness Ratio (HR)}

%I asked participants choose if they wanted to help us train my ECA. If they chose to help, they needed to write a script in the questionnaire. Participants imagined a conversation between themselves and the ECA using a template (Table 3) to write the content of the conversation. 
The portion of participants who chose to help train the ECA (HP) were 71\% (n=55) in the laughing group, 72\% (n=50) in the smiling group, and 72\% (n=73) in the neutral group (Figure~\autoref{fig:results-hp}). A Chi-square test indicated no statistically significant differences, \emph{p} = .966, meaning that willingness to help did not appear to be affected by ECA version. A Chi-square test indicated that the same was true for novices (helped n=152, declined n=6) compared to non-novices (helped n=66, declined n=5), \emph{p} = .102 (Figure~\autoref{fig:results-hpratio}). However, Mann-Whitney U tests found statistically significant relationships between GTA and HP, \emph{U} = 7816.5, \emph{Z} = 3.42, \emph{p} < .001, and GAIA and HP, \emph{U} = 8034.5, \emph{Z} = 3.85, \emph{p} < .001.

For those who chose to help (n = 178, 71\%), a Kruskall-Wallis test found no statistically difference for HR between the laughing group (M = 49.5, SD = 
30.7), smiling group (M = 51.0, SD = 28.0), or neutral group (M = 46.1, SD = 34.3), \emph{p} = .396 (Figure~\autoref{fig:results-hr}). Nor did a Mann-Whitney U test find a statistically significant difference between novices (M = 48.1, SD = 31.7) and non-novices (M = 50.7, SD = 30.4), \emph{p} = .685. However, a Spearman's rank correlation test for GAT and HR was statistically significant, \emph{r}(180) = .314, \emph{p} < .001, as was one for GAIA and HR, \emph{r}(180) = .376, \emph{p} < .001).
%the degree of helpfulness as measured by counts of words (HR) was M=48.9 for Laughing group, M=46.9 for smiling group, and M=45.5 for the neutral group. A Kruskal-Wallis H test indicated that there was no statistically significant difference by group, \emph{χ2} = .97, \emph{p} = .615. %, with a mean rank score of 113.58 for the Laughing group, 110.98 for the Smiling group, 104.01 for the Neutral group, which means that there was no significant difference between different groups for HP. There was no significant difference in participants' contribution to the three groups, 

In short, the differing emotional expressions of the ECA had no real impact on willingness to help or degree of aid, and neither did familiarity. Instead, general technology and IA acceptance seems to have had a greater impact.

\subsection{Results Summary and H1}
Altogether, the results fail to support H1. The use of laughter and humour as well as smiling had no effect on initial acceptance of the ECA. Instead, general technology acceptance seemed to explain patterns in acceptance of the ECA.

%% ----------------- DISCUSSION -----------------
\section{Discussion}

The aim of this research was to investigate the influence of humour, laughter, and positive expressions in an ECA as a facet of first impressions that may influence older adult's initial acceptance of the ECA. %This study focused on comparing the general acceptance of technology (GAT) and general acceptance of intelligent agents (GAIA) among older adults while also examining the impact of incorporating positive initial engagement in ECAs. Results indicated that there was no statistically significant difference between the Laughing, Smiling, and Neutral groups regarding the acceptance of the ECA (KONO). 
We employed a rigorous theoretical and design process, including multiple pilot tests and manipulation checks. Yet, the expected result was not found: humour and laughter, as well as general positivity expressed through verbal puns and nonverbal smiles, %, expressed in the ECA's nonverbal and verbal behaviour 
had no apparent effect. This included attitudes as well as behavioural measures. 
We could, however, explain this result through general attitudes towards technology and IA. A notable exception was the hint of a difference between novices and non-novices, but this did not bear out. In short, humour could not improve acceptance in the context of first impressions with the ECA, nor could ECA positivity, at least in the case of our ``Mikan.'' 

Prior research has emphasized the significance of positive emotions and dynamic features in ECAs for older adults~\cite{nijholt2002embodied,ring2013addressing}. %Both studies [64, 71] emphasize the importance of incorporating emotionally expressive and engaging features in ECAs to enhance the user experience and increase acceptance among older adults. 
For instance, Ring at el.~\cite{ring2013addressing} suggested that humour might have a more substantial impact on user acceptance than what we found. Our results may be attributed to differences in ECA design~\cite{seeger2021texting}, the cultural backgrounds of participants~\cite{anderson2017technology}, variations in research methods across studies~\cite{seeger2021texting}, and how older adults approach humour, i.e., individual and personality factors~\cite{Leist2012,Martin2003}. Even age may play a role: the type of humour that older adults find most enjoyable may differ from other forms~\cite{mallya2019theoretical}. %Moreover, while humour may not directly affect acceptance, it may promote engagement in other ways and ensure a positive experience. 
%One possible explanation for the differences in the impact of humour on acceptance could be related to how humour was implemented in the ECA. In particular, the 
Differences could also relate to the specific type of humour selected or the specific way it was incorporated in the ECA, e.g., the choice of joke and particular facial expressions. Humour preferences may vary among individuals: what may be humorous to one person may not be to another.

Future work can explore these possibilities. For example, ECAs can be designed to adapt their humour styles based on users' personality traits to create more personalized and engaging interactions. Additionally, ECAs can be programmed to recognize and respond to the emotional state of a user, allowing for the strategic use of humour to elevate the mood of a user and enhance their overall experience with the technology. Personality traits and mental health can also be used as variables in ECA acceptance studies and can provide valuable insights for developing targeted interventions. For example, if humour was found to have a more significant impact on individuals with specific personality traits or mental health, then interventions could be designed to use humour to increase ECA acceptance among these groups.

%Add discussion on “differences in expectation about the ECA's functionality also influence how humor might impact their perceived acceptance?” given how we framed the ECA for participants in a rather broad way (“personal assistant” who does a million things)
How the ECA was framed as a general personal assistant with a wide array of ever-customizable interests could also have influenced responses in general and especially towards its expressions of humour and positivity. Mikan introduced ``herself'' and ``her'' functionalities in a certain way: able to read, converse, buy things, and even play baseball---plus whatever else the user was interested in. This is a rather broad framing of mostly transactional abilities that may not lend itself to humour. Future work should explore what ECA ``types'' may be best and least suited to humour.

Humour can also be influenced by cultural factors, as Easterners and Westerners often have different attitudes towards humour~\cite{jiang2019cultural}. As such, perceptions of humour and its impact on the acceptance of ECA may be influenced by cultural background. We should therefore design humour in accordance with different cultures. Therefore, the incorporation of humour in ECAs designed for Japanese users should be culturally sensitive and consistent with current cultural norms and preferences. humour that resonates positively with Japanese participants can increase their engagement with the ECA and promote more enjoyable and natural interactions. 
%To deepen the discussion, it would be valuable to compare the findings of this study with those of other cultural contexts. Such a comparative analysis could reveal whether the effect of humour on ECA acceptance holds across cultural contexts or whether there are differences in the perception and acceptance of ECA based on cultural contexts. In summary, recognizing the influence of cultural factors on humour and technology acceptance is critical to designing ECAs that are appropriate for a given cultural context and that resonate with users. Understanding these cultural nuances can help develop more effective and engaging ECAs, ultimately improving user acceptance and user experience, especially among older adults in Japan.

While little work exists, some suggests that older adults may face challenges when it comes to accepting humour expressed in technology~\cite{Monahan2014}. O'Connell et al.~\cite{OConnell2021} indicated that lack of exposure to technology creates an additional psychological barrier to adoption of new technology. In our case, most participants had little or no prior experience with ECAs, making it their first encounter with this advanced technology. Consequently, older adults unacquainted with the intricacies of ECAs may display hesitation, even when more ``humanlike'' factors, such as humour, are integrated into the design. %Despite positive initial engagement with the ECAs, it became evident that this alone was insufficient to significantly increase the older adults' acceptance of technology. This suggests that 
An approach tailored to the specific desires and barriers faced by older adults may be crucial for enhancing their receptiveness to ECAs. Perhaps first impressions are not enough. Future work may explore longer-term engagements or run follow-up studies wherein older adults directly interact with the ECA to further investigate the relationship between humour and acceptance.

\subsection{Limitations}
We acknowledge several limitations in this work, particularly with the online setup. Online experiments cannot guarantee high quality due to various uncontrollable factors (e.g., environment setup and sudden distractions). While we used an attention check, we cannot confirm that participants were fully attentive (e.g., savvy respondents could have skip-searched the video for the code, although we tried to account for this shortcut by removing respondent data less than 6 minutes). %External interferences may also have affected responses.
There were also uneven distributions in terms of gender (over-representation of men) and ECA group (due to the randomization procedure and elimination of data after quality checks). This may have introduced bias into the results, as participant gender may relate to ECA acceptance~\cite{Chen2014}, and we used a feminine ECA. While we avoided stereotype threats~\cite{spencer2016stereotype} by asking about demographics at the end of the study, future work should aim for accurate gender representation. 

Participants did not directly interact with the ECA. Since we were focused on first impressions, we believe that our setup reflected a typical introductory experience. Still, the ``engagement'' was shortand without a follow-up interactive experience. Future work should consider first impressions in longer-term engagements.

%R3: Add discussion/limitation: “For example, would people have been more willing to share personal information if the agent is seen as more personable, light-hearted and affable? How did the context of the conversation (I.e., the script and conversation) and requirements imposed on the user (e.g., tasks or responses that are requested by the CA of the user) impact the outcome, and could the result potentially differ under other contexts?”
%R3: Add limitation: “Would it not have been better for participants to reply as if they were talking directly to the agent themselves, rather than scripting an interaction with the agent? (This would then have simulated a direct interaction with the agent as well, and would have not relied on hypothetical that, while used in prior work at CHI, could be seen as less rigorous by some readers).”
We used a novel measure of helpfulness towards the ECA. While we based it on previous, similar research, other features of the research design could have affected results. For instance, participants may have been more open to helping the ECA after an interactive experience, especially a conversational one, expressions of keen interest in the user, and/or if given a lighter task. Participants could have also been given the task indirectly, i.e., asked to converse through text input with the ECA. This would require some automation in the online study environment. Future work may explore these theoretical and technical possibilities. The development and validation of standard objective, behavioural measures would be a boon for continued online experiments and meta-analysis work.

%Furthermore, the use of the "Wizard of Oz" method introduces another limitation. This approach involves human operators providing something to participants, which may introduce variability and potential confounding factors. In this research, a lab member controlled the facial expression of the ECA, so participants may have felt it was unnatural. Also, a feminine ECA was used, which may influence the results of the experiment in complex ways that future research should consider directly in the study design [43].

%Lastly, I also think that different responses were given to participants based on their personalities which may introduce confounding variables and affect the reliability of the results [63, 70]. In future research, I may like to research about how different personalities of participants influence the acceptance for the ECAs.

%% ----------------- CONCLUSION -----------------
\section{Conclusion}
%This study contributes to understanding the complex relationship between positive initial engagement and ECA acceptance with older adults. 
Positive first impressions and humour did not impact older adults' initial acceptance of an ECA. %Still, a promising design element based on previous research that can potentially create engaging interactions and positive emotional experiences for ECAs. All in all, 
We should deeply explore the nuances of humour in ECAs for older adults to identify the most effective and culturally sensitive approaches. %Understanding cultural context and preferences will be crucial in successfully implementing humour in ECAs and enhancing their acceptance among older adults. 
Even if humour itself does not directly affect acceptance, it may play a role in longer-term engagements. %factors other than initial engagement play a critical role in determining acceptance among older adults. 
Designers may take a holistic and longitudinal approach using a range of other factors, such as usability, functionality, familiarity with the technology, and personal preferences, perhaps by asking older adults for their favourite jokes. %Understanding these multiple factors may help create more effective, user-centered ECAs that meet the unique needs and preferences of older adults, ultimately increasing their acceptance and usability. By considering a variety of design factors and accommodating individual preferences, designers can develop ECAs that are better suited to the needs of older adults, ultimately improving their overall health and the quality of their interactions with technology.

\begin{acks}
This work was supported by departmental funds. We thank the members of the Aspirational Computing Lab for pilot testing and support, especially Suzuka Yoshida. We thank Xiuzhu Gu and Hiroyuki Umemuro for early feedback. Katie Seaborn conscientiously dissents to in-person participation at CHI '24 this year; read their positionality statement here: \url{https://bit.ly/chi24statement}
\end{acks}

\balance

\bibliographystyle{ACM-Reference-Format}
\bibliography{REFS}

\end{CJK}
\end{document}